\documentclass[aps,pra,twocolumn,showpacs,superscriptaddress,nofootinbib]{revtex4-1}
\usepackage{graphicx}
\usepackage{color}
\usepackage{transparent}
\usepackage{amsthm,amsmath,amssymb,amsfonts,amssymb}
\usepackage{float}
\usepackage{hyperref}
\usepackage{tikz}	
\usetikzlibrary{backgrounds,fit,decorations.pathreplacing,decorations.pathmorphing}

\newcommand{\bra}[1]{\ensuremath{\left\langle#1\right|}}
\newcommand{\ket}[1]{\ensuremath{\left|#1\right\rangle}}

\begin{document}

\title[]{Symmetry in the open-system dynamics of quantum correlations}
\author{Henri Lyyra}
\email{hesaly@utu.fi}
\affiliation{Turku Center for Quantum Physics, Department of Physics and Astronomy, University of Turku, FIN-20014, Turun yliopisto, Finland}
\author{G\"{o}ktu\u{g} Karpat}
\affiliation{Faculdade de Ci\^encias, UNESP - Universidade Estadual Paulista, Bauru, SP, 17033-360, Brazil}
\author{Chuan-Feng Li}
\affiliation{Key Laboratory of Quantum Information, University of Science and Technology of China, CAS, Hefei, 230026, People’s Republic of China}
\affiliation{Synergetic Innovation Center of Quantum Information and Quantum Physics, University of Science and Technology of China, Hefei, 230026, People’s Republic of China}
\author{Guang-Can Guo}
\affiliation{Key Laboratory of Quantum Information, University of Science and Technology of China, CAS, Hefei, 230026, People’s Republic of China}
\affiliation{Synergetic Innovation Center of Quantum Information and Quantum Physics, University of Science and Technology of China, Hefei, 230026, People’s Republic of China}
\author{Jyrki Piilo}
\affiliation{Turku Center for Quantum Physics, Department of Physics and Astronomy, University of Turku, FIN-20014, Turun yliopisto, Finland}
\author{Sabrina Maniscalco}
\affiliation{Turku Center for Quantum Physics, Department of Physics and Astronomy, University of Turku, FIN-20014, Turun yliopisto, Finland}
\affiliation{Centre for Quantum Engineering, Department of Applied Physics,  School of Science, Aalto University, P.O. Box 11000, FIN-00076 Aalto, Finland}

\begin{abstract}
We study the symmetry properties in the dynamics of quantum correlations for two-qubit systems in one-sided noisy channels, with respect to a switch in the location of noise from one qubit to the other. We consider four different channel types, namely depolarizing, amplitude damping, bit-flip, and bit-phase-flip channel, and identify the classes of initial states leading to symmetric decay of entanglement, non-locality and discord. Our results show that the symmetric decay of quantum correlations is not directly linked to the presence or absence of symmetry in the initial state, while it does depend on the type of correlation considered as well as on the type noise. We prove that asymmetric decay can be used to infer, in certain cases, characteristic properties of the channel. We also show that the location of noise may lead to dramatic changes in the persistence of phenomena such as entanglement sudden death and time-invariant discord.
\end{abstract}

\maketitle

\section{Introduction}

Correlations of genuine quantum nature among the individual constituents of composite systems play a fundamental role in quantum physics.  Entanglement is the paramount example of quantum correlations, considered by the founding fathers of quantum physics as the most bizarre aspect of this theory  \cite{schro}. Initially viewed as a mere philosophical subject, it gained popularity and importance with the development of quantum information theory, when it was recognized as a resource for several tasks such as teleportation, superdense coding, and quantum key distribution \cite{nielsen,entreview}.

A specific type of quantum correlation possessed by some entangled states is associated to the concept of quantum non-locality, which implies that predictions of quantum mechanics cannot be simulated by a local hidden variable model \cite{bell,gisin,werner}. The presence of non-local correlations in bipartite quantum systems leads to violation of Bell-type inequalities, such as the Clauser-Horne-Shimony-Holt (CHSH) inequality \cite{CHSH}. States violating Bell inequalities are crucial for certain quantum technologies such as secure quantum communication \cite{key}.

For mixed states, there exists a broader type of quantum correlations that does not occur in classical systems, namely quantum discord. It has been demonstrated that discordant states can perform more efficiently than their classical counterparts in certain applications \cite{sep}. As a consequence, numerous different measures of discord have been introduced in the recent literature to characterize quantum correlations more general than entanglement \cite{discord1, discord2, discord3, discord4, discorduse1, discord5, discord6}.

Like all crucially quantum properties, entanglement, discord and non-locality are fragile and generally quickly disappear in presence of noise induced by the environment. Discord is clearly more robust than entanglement to the effects of noise, while entanglement may disappear after a finite time (sudden death of entanglement), discord decays asymptotically \cite{werlang}. Moreover, for certain types of local noise discord may remain constant in time, although the state of the system evolves (time-invariant discord) \cite{haikka, addis}.

In this paper we investigate the dynamics of quantum correlations, namely discord, entanglement and non-locality, for two-qubit systems subjected to various types of one-sided noisy channels. More specifically we focus on a class of states known as X states for which all quantum correlations can be calculated analytically, and we study their symmetry properties with respect to a change in the location of the noise from one qubit to the other. Asymmetric decay of entanglement was earlier studied briefly in \cite{hdecki}. Changing the noise location corresponds, for example, to a situation in which Alice produces a state possessing certain quantum correlations and sends it to Bob, who measures the received state. The two parts of the quantum correlated state travel  along different paths of equal length, named $U$ (upper) and $L$ (lower). An eavesdropper Eve may attack either one path or the other, her attack being modeled by introducing noise of different type (depolarizing, amplitude damping, bit-flip, and bit-phase-flip channel). 

We are interested in understanding whether the dynamics of various quantum correlations is sensitive, and if so how much, to the location of Eve's attack, more precisely to whether she eavesdrops along the $U$ or $L$ path. Of course, we expect that the answer to this question will depend on both the initial quantum correlated state that Alice prepares, and the type of noise introduced by Eve. More precisely, we are interested in identifying the classes of initial states leading to symmetric behavior, with respect to noise acting on either $U$ or $L$, for the different types of channels and for different types of quantum correlations. Moreover, we investigate how such classes of initial states  change for different types of quantum correlations and if there are overlaps between the classes of initial states that lead to symmetric behaviour of quantum entanglement, non-locality and discord.

More specifically, our study gives an answer to the following interesting questions:

1) Is there any connection between symmetry properties of the initial state and symmetry properties of (i) the state dynamics and (ii) the dynamics of quantum correlations (discord, entanglement, non-locality) with respect to a switch in the noise between the channel? In other words, are certain symmetries in the initial state necessary to guarantee a symmetric dynamics of quantum correlations?

2) How sensitive are effects such as entanglement sudden death and time-invariant discord to the location of noise (along the $U$ or $L$ path)? Stated another way, are there situations for which entanglement sudden death or time-invariant discord occurs only when Eve attacks along the $U (L)$ path but not when she attacks the $L (U)$ path, all other conditions being the same?

3) Assuming that we do not know which type of noise acts on either $U$ or $L$, can Bob use the symmetric/asymmetric decay properties of quantum correlations to infer or characterize the type of noise, under minimal assumptions?

Our paper is structured as follows. In Sec. II we introduce the class of initial states considered and discuss the symmetry properties of the dynamics and of entanglement as measured by concurrence. In Sec. III we analyze some physical consequences of our findings such as the effect of noise location on entanglement sudden death, the use of asymmetric decay for channel discrimination, and the connection between entanglement decay and entropy. In Sec. IV we discuss the symmetry properties of non-locality and compare the classes of initial states leading to symmetric or asymmetric decay for entanglement and non-locality, respectively. In Section V we discuss the symmetry properties of a distance-based measure of quantum discord and the effect of the noise location on time-invariant discord. Finally in Sec. VI we summarize and present conclusions.

\section{X states, concurrence and one-sided channels}\label{concurr}

X states are a subclass of 2-qubit states appearing naturally in physical processes \cite{bose, pratt, mdms, wang}. We denote an arbitrary X state by $X$, where
\begin{align}\label{xstat}
X&
=
\begin{pmatrix}
\rho_{11}		&0				&0				&\rho_{14}	\\
0				&\rho_{22}	&\rho_{23}	&0	\\
0				&\rho_{23}^*	&\rho_{33}	&0	\\
\rho_{14}^*	&0				&0				&\rho_{44}	
\end{pmatrix},
\end{align}
$\rho_{11},\, \rho_{22},\, \rho_{33},\, \rho_{44} \in \mathbb{R}$, $\rho_{11} + \rho_{22} + \rho_{33} + \rho_{44} =1$, $\rho_{23},\,  \rho_{14} \in \mathbb{C}$,  \color{black} and the density matrix is written in the basis $\mathcal{B}:=\{\ket{00},\,\ket{01},\,\ket{10},\,\ket{11}\}$\color{black}. \color{black}Generally, w\color{black}e say, that \textit{a state is swap symmetric} if the corresponding density matrix is invariant under  swapping the elements according to the rule $\ket{ij}\rightarrow \ket{ji}$. A straightforward calculation shows that $X$ is symmetric if and only if the following conditions are satisfied:
\begin{equation}
\begin{aligned}\label{symm1}
\rho_{22}\,  &= \rho_{33}\,, \\ 
\rho_{23} &\in \mathbb{R}\,.
\end{aligned}
\end{equation}
The first condition is equivalent to  $\text{tr}[\sigma_3^UX] = \text{tr}[\sigma_3^LX]$ and the second one to $\text{tr}[\sigma_1^U\otimes\sigma_2^LX] = \text{tr}[\sigma_2^U\otimes\sigma_1^LX]$, where $\sigma_1,\,\sigma_2,\,\, \text{and}\,\,\sigma_3$ are the Pauli matrices. Here superscripts $U$ and $L$ refer to operators on Hilbert spaces $\mathcal{H}^U$ and $\mathcal{H}^L$ of qubits $U$ and $L$, respectively. In the vector notation of basis $\mathcal{B}$, the first qubit corresponds to $U$ and the latter one to $L$. The conditions in Eq.~\eqref{symm1} are satisfied for instance by all Bell-diagonal and Werner states, which form two important subclasses of X states. 

\color{black}Since entanglement is an essential feature of quantum mechanics, quantifying it has been an active field of research. Multiple entanglement measures have been defined, the most popular of which is \textit{concurrence} \cite{wootters}. \color{black} 
It has been shown \cite{xstateconcurrence} that, for all X states, concurrence can be obtained directly from the matrix elements as:
\begin{equation}\label{conc}
C(X) = 2\max\big{\{}0,\vert \rho_{23}\vert - \sqrt{\rho_{11}\rho_{44}}, \vert \rho_{14}\vert - \sqrt{\rho_{22}\rho_{33}}\big{\}}.
\end{equation} 

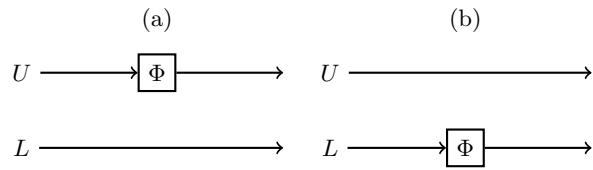
\begin{figure}[t]
  \centerline{
    \begin{tikzpicture}[thick]
    \tikzstyle{operator} = [draw,fill=white,minimum size=1.5em] 
    \tikzstyle{phase} = [fill,shape=circle,minimum size=5pt,inner sep=0pt]
    \tikzstyle{surround} = [fill=blue!10,thick,draw=black,rounded corners=2mm]
        \node at (-2.3,2.2) (qb) {(a)};
        \node at (-4.1,1.5) (q2a) {$U$};
    \node at (-4.1,0.5) (q3a) {$L$};
    \node[operator] (q4a) at (-2.3,1.5) {$\Phi$} edge [<-] (q2a);
    \node[-] (q10a) at (-0.5,1.5) {} edge [<-] (q4a);
    \node[-] (q20a) at (-0.5,0.5) {} edge [<-] (q3a);
        \node at (1.8,2.2) (qd) {(b)};
        \node at (0,1.5) (q2b) {$U$};
    \node at (0,0.5) (q3b) {$L$};
    \node[operator] (q4b) at (1.8,0.5) {$\Phi$} edge [<-] (q3b);
    \node[-] (q10b) at (3.6,0.5) {} edge [<-] (q4b);
    \node[-] (q20b) at (3.6,1.5) {} edge [<-] (q2b);
    \end{tikzpicture}
  }
  \caption{
Visualization of different noise locations. Alice sends a two-qubit system to Bob. Qubits $U$ and $L$ are transmitted through the upper and lower path, respectively.  In (a) and (b) the local noise noise influences the qubit $U$ and $L$, respectively. 
}
\label{ul}
\end{figure}

Physical dynamics of a quantum state $\rho$ is represented by completely positive and trace preserving linear maps called channels. A map $\Phi$ 
 is a channel if and only if there exists such set of operators $\{K_i\}_{i\in \mathcal{I}}$ that
\begin{align}\label{kraus}
&
\Phi(\rho)=\sum_{i\in \mathcal{I}} K_i \rho K_i^{\dagger}\,,\,\,\,\,\,
\sum_{i\in \mathcal{I}} K_i^{\dagger}K_i = I\,,
\end{align}
for all states $\rho$ \cite{cp}. The decomposition in Eq.~\eqref{kraus} is called the Kraus decomposition and the operators $K_i$ are called the Kraus operators.

In this study we concentrate on  one-sided channels, which means that the channel influences only one of the local qubit states at a time, see Fig.~\ref{ul}. This means that the Kraus operators acting on $\mathcal{H}^U\otimes \mathcal{H}^L$ are of the form $K_i:=K_i'\otimes I^L$, when the channel affects qubit $U$ and of the form $K_i:=I^U\otimes K_i'$ when the channel affects qubit $L$. Here $I^U$ and $I^L$ are the identity operators of $\mathcal{H}^U$ and $\mathcal{H}^L$, respectively. We denote the one-sided channels affecting qubits $U$ and $L$ by
\begin{align}\label{totkrausa}
\Phi^U(\rho) &:= \sum_{i\in \mathcal{I}} (K_i'\otimes I^L) \rho (K_i'^{\dagger}\otimes I^L)\,,\\ \label{totkrausb}
\Phi^L(\rho) &:= \sum_{i\in \mathcal{I}} (I^U\otimes K_i') \rho (I^U\otimes K_i'^{\dagger})\,,
\end{align}
respectively. We say that \textit{the dynamics of a state $\rho$ is symmetric} if $\Phi^U(\rho) = \Phi^L(\rho)$.
Next we study the conditions leading to symmetric and asymmetric state dynamics and entanglement decay under the effects of different channels\footnote{\color{black}For pure dephasing channel, given by Kraus operators $K_1' = \sqrt{1-p}I,\,\sqrt{p}\sigma_3$, we notice that $\Phi_p^U(X) = \Phi_p^L(X)$ for all initial X states $X$. Thus all the properties of the system, such as entanglement, Bell function and quantum discord, evolve symmetrically with respect to the location of the noise. \color{black}}.
\subsection{Depolarizing channel}\label{dep}
\color{black}Corresponding to Eq.~\eqref{totkrausa} and \eqref{totkrausb}, the dynamics of an arbitrary X state under one-sided depolarizing channels can be written as\color{black}
\begin{align}\nonumber
&\Phi_p^U(X)=\\&
\begin{pmatrix}
\rho_{11} + \frac{p}{2}r_{31}	&0							&0						&q\rho_{14}	\\
0						&\rho_{22} + \frac{p}{2}r_{42} 	&q\rho_{23}				&0	\\
0						&q\rho_{23}^*			&\rho_{33} + \frac{p}{2}r_{13}	&0	\\
q\rho_{14}^*				&0							&0							&\rho_{44} +\frac{p}{2}r_{24}
\end{pmatrix}\,,\label{dephcana}
\\
&\Phi_p^L(X)=\nonumber
\\&
\begin{pmatrix}
\rho_{11} + \frac{p}{2}r_{21}	&0							&0						&q\rho_{14}	\\
0						&\rho_{22} + \frac{p}{2}r_{12} 	&q\rho_{23}				&0	\\
0						&q\rho_{23}^*			&\rho_{33} + \frac{p}{2}r_{43}	&0	\\
q\rho_{14}^*				&0							&0							&\rho_{44} +\frac{p}{2}r_{34}
\end{pmatrix}\,,
\label{dephcanb}
\end{align}
\color{black}where $p\in[0,1]$ is the channel strength parameter, telling how strongly the channel influences states and we have denoted $r_{jk}:=\rho_{jj}-\rho_{kk},\,q:=1-p$.
\color{black}
By comparing Eq.~\eqref{dephcana} and \eqref{dephcanb}, it is evident, that the dynamics of a state is symmetric if and only if $\rho_{11} = \rho_{44}$ and $\rho_{22} = \rho_{33}$. We note that the symmetry of the state dynamics  is independent of the phase of $\rho_{23}$, unlike the symmetry of the initial state, but it requires $\rho_{11} = \rho_{44}$ instead. Trivially, symmetric density matrix dynamics implies symmetric behavior of all  system properties and thus leads to symmetric entanglement decay. By using Eq.~\eqref{conc} we get the concurrences of the output states as
\begin{equation}
\begin{aligned}
&C\textbf{(}\Phi_p^U(X)\textbf{)}=\\
&2\max\Big{\{}0, 
q\vert \rho_{23}\vert - \frac{1}{2}\sqrt{(2\rho_{11} + r_{31}p)(2\rho_{44} + r_{24}p)}
,\\ \label{depca}
& q\vert \rho_{14}\vert - \frac{1}{2}\sqrt{(2\rho_{33} + r_{13}p)(2\rho_{22} + r_{42}p)}\Big{\}},
\end{aligned}
\end{equation}
and
\begin{equation}
\begin{aligned}
&C\textbf{(}\Phi_p^L(X)\textbf{)}=\\ 
&2\max\Big{\{}0,
q\vert \rho_{23}\vert - \frac{1}{2}\sqrt{(2\rho_{11} + r_{21}p)(2\rho_{44} + r_{34}p)},\\ \label{depcb}
& q\vert \rho_{14}\vert - \frac{1}{2}\sqrt{(2\rho_{22} + r_{12}p)(2\rho_{33} + r_{43}p)}\Big{\}}.
\end{aligned}
\end{equation}
A straightforward calculation shows that entanglement decays symmetrically, i.e. $C\textbf{(}\Phi_p^U(X)\textbf{)} = C\textbf{(}\Phi_p^L(X)\textbf{)}\, \forall p\in[0,1]$, if and only if $\rho_{33} = \rho_{22}$ or $\rho_{11} = \rho_{44}$. The first condition is necessary for the symmetry of the initial state, as formulated in Eq.~\eqref{symm1}. Instead the second one, $\rho_{11} = \rho_{44}$, 
is not.

We conclude that symmetry of entanglement decay requires neither swap symmetry of the initial state nor symmetry in the dynamics of the state.
\subsection{Amplitude damping channel}\label{amp}
\color{black}
The dynamics of an arbitrary X state under one-sided amplitude damping channels can be written as\color{black}
\begin{align}
\Phi_p^U(X)&=
\begin{pmatrix}
\rho_{11} + p\rho_{33}				&0					&0						&\sqrt{q}\rho_{14}	\\
0					&\rho_{22} + p\rho_{44}			&\sqrt{q}\rho_{23}			&0	\\
0					&\sqrt{q}\rho_{23}^*	&q\rho_{33}				&0	\\
\sqrt{q}\rho_{14}^*	&0					&0						&q\rho_{44}
\end{pmatrix},\label{ampcana}
\\
\Phi_p^L(X)&=
\begin{pmatrix}
\rho_{11} + p\rho_{22}				&0					&0						&\sqrt{q}\rho_{14}	\\
0					&q\rho_{22}			&\sqrt{q}\rho_{23}			&0	\\
0					&\sqrt{q}\rho_{23}^*	&\rho_{33} + p\rho_{44}				&0	\\
\sqrt{q}\rho_{14}^*	&0					&0						&q\rho_{44}
\end{pmatrix}\label{ampcanb}
.
\end{align}
Comparison of Eq.~\eqref{ampcana} and \eqref{ampcanb} shows, that the dynamics of a state is symmetric if and only if $X = \ket{11}\bra{11}$, which is invariant. Since for this state $\rho_{22} = \rho_{33} = \rho_{23} = 0$, symmetric state dynamics occurs only for a single state which is swap symmetric, unlike in the case of depolarizing channel. 
By \eqref{conc}, we get the concurrences of the output states as
\begin{equation}
\begin{aligned}\label{ampca}
C\textbf{(}\Phi_p^U(X)\textbf{)}=2\max\Big{\{}0,&\sqrt{q}\Big(\vert \rho_{23}\vert - \sqrt{\rho_{44}(\rho_{11} + \rho_{33}p)}\Big)
,\\ 
& 
\sqrt{q}\Big(\vert \rho_{14}\vert - \sqrt{\rho_{33}(\rho_{22} + \rho_{44}p)}\Big)\Big{\}}, 
\end{aligned}
\end{equation}
\begin{equation}
\begin{aligned}
\label{ampcb}
C\textbf{(}\Phi_p^L(X)\textbf{)}=2\max\Big{\{}0,&\sqrt{q}\Big(\vert \rho_{23}\vert - \sqrt{\rho_{44}(\rho_{11} + \rho_{22}p)}\Big), \\  
& 
\sqrt{q}\Big(\vert \rho_{14}\vert - \sqrt{\rho_{22}(\rho_{33} + \rho_{44}p)}\Big)\Big{\}}.
\end{aligned}
\end{equation}
Now entanglement decay is symmetric if and only if $\rho_{33} = \rho_{22}$ or $\rho_{44} = 0$. The first condition is necessary for the symmetry of the initial state but the second one, $\rho_{44} = 0$, is not related to it. 

So again, an initially asymmetric state can lead to symmetric decay of entanglement. Also, initial states leading to asymmetric state dynamics can have symmetric decay for entanglement.
%
\begin{figure}[t]
  \centerline{
    \begin{tikzpicture}[thick]
    \tikzstyle{operator} = [draw,fill=white,minimum size=1.5em] 
    \tikzstyle{phase} = [fill,shape=circle,minimum size=5pt,inner sep=0pt]
    \tikzstyle{surround} = [fill=blue!10,thick,draw=black,rounded corners=2mm]
        \node at (-2.3,2.2) (qb) {(a)};
        \node at (-4.1,1.5) (q2a) {$U$};
    \node at (-4.1,0.5) (q3a) {$L$};
    \node[operator] (q4a) at (-2.3,1.5) {$\Phi_p$} edge [<-] (q2a);
    \node[operator] (q5a) at (-2.3,0.5) {$\Psi_p$} edge [<-] (q3a);
    \node[-] (q10a) at (-0.5,1.5) {} edge [<-] (q4a);
    \node[-] (q20a) at (-0.5,0.5) {} edge [<-] (q5a);
        \node at (1.8,2.2) (qb) {(b)};
    \node at (0,1.5) (q2b) {$U$};
    \node at (0,0.5) (q3b) {$L$};
    \node[operator] (op21b) at (1.3,1.5) {$\Phi_p$} edge [<-] (q2b);
    \node[operator] (q4b) at (2.35,1.5) {$\Psi_p$} edge [-] (op21b);
    %
    %
    \node[-] (q10b) at (3.6,1.5) {} edge [<-] (q4b);
    \node[-] (q20b) at (3.6,0.5) {} edge [<-] (q3b);
        \node at (-2.3,-0.3) (qc) {(c)};
    \node at (-4.1,-1) (q2c) {$U$};
    \node at (-4.1,-2) (q3c) {$L$};
    \node[operator] (op21c) at (-2.8,-1) {$\Phi_p$} edge [<-] (q2c);
    \node[operator] (q4c) at (-1.75,-1) {$\Psi_p$} edge [-] (op21c);
    \node[operator] (q5c) at (-2.3,-2) {$\Xi_p$} edge [<-] (q3c);
    \node[-] (q10c) at (-0.5,-1) {} edge [<-] (q4c);
    \node[-] (q20c) at (-0.5,-2) {} edge [<-] (q5c);
        \node at (1.8,-0.3) (qd) {(d)};
    \node at (0,-1) (q2d) {$U$};
    \node at (0,-2) (q3d) {$L$};
    \node[operator] (op21d) at (0.9,-1) {$\Phi_p$} edge [<-] (q2d);
    \node[operator] (op22d) at (1.8,-1) {$\Psi_p$} edge [-] (op21d);
    \node[operator] (q4d) at (2.7,-1) {$\Xi_p$} edge [-] (op22d);
    \node[-] (q10d) at (3.6,-1) {} edge [<-] (q4d);
    \node[-] (q20d) at (3.6,-2) {} edge [<-] (q3d);
    \end{tikzpicture}
  }
  \caption{
Schematic illustrations (a), (b), (c), and (d) visualize four different geometric  configurations for noise combinations. Here channels $\Phi_p,\,\Psi_p,$ and $\Xi_p$ correspond to local amplitude damping or depolarizing noises with equal channel strength parameters $p$ acting on qubits $U$ and $L$.
}
\label{combi}
\end{figure}
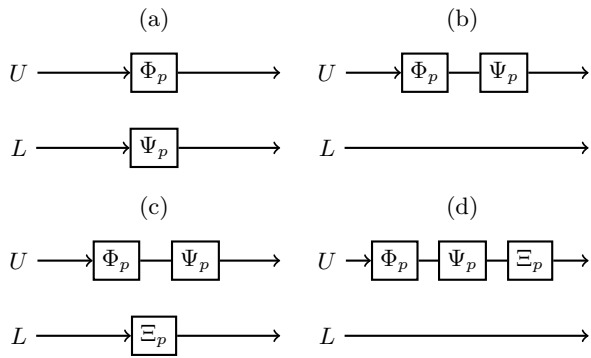
\subsection{Entanglement decay in channel combinations}\label{comb}
 To generalize the analysis, we study the initial conditions for entanglement decay also in channel combinations. 
 By combining  local  amplitude damping and depolarizing channels with equal channel strengths $p$, we can create new channels for the two-qubit system. In Fig.~\ref{combi} we illustrate the most simple two channel combinations. We have solved the families of initial states leading to symmetric and asymmetric entanglement decay in each noise configuration (a)--(d) with the same analysis as used in Sec.~\ref{dep}--\ref{amp}.

Trivially, entanglement decay is symmetric for all initial states in configuration (a) if $\Phi = \Psi$. If in configuration (b) the channels are chosen as $\Phi = \Psi$, the conditions for symmetric entanglement decay are the same as in the situation of channel $\Phi$ influencing just one of the qubits once. The same applies also for configurations (c) and (d): whenever $\Phi = \Psi = \Xi$, the conditions for symmetric entanglement decay are the same as in the case of $\Phi$ affecting just one of the qubits once. This means, that in the sense of entanglement decay symmetry, adding identical copies of the same channel does not break the symmetry or create it.

On the other hand, if $\Phi\neq\Psi$ in (a), entanglement decay is symmetric if and only if $\rho_{22} = \rho_{33}$. Also, if two of the channels in (c) are different, entanglement decays symmetrically if and only if $\rho_{22} = \rho_{33}$. If in (b) and (d) two of the channels are different, entanglement decays symmetrically if and only if $\rho_{22} = \rho_{33}$ or $\rho_{11} = \rho_{44} = 0$. First of these is just one of the swap symmetry conditions, leading trivially to symmetric entanglement decay but the second one is actually a condition which satisfies the non-trivial symmetry conditions of both depolarizing and amplitude damping channels, presented in Sec.~\ref{dep}--\ref{amp}.

\subsection{Bit-flip and bit-phase-flip channels}\label{bfbpf}
To avoid redundancy, we present here only the results for bit-flip and bit-phase-flip channels. For further details we refer the reader to appendices \ref{bf}--\ref{bpf}. 

We see that, for bit-flip channel, the dynamics of a state is symmetric if and only if $\rho_{11} = \rho_{44},\,\rho_{22} = \rho_{33}$ and $\rho_{14},\, \rho_{23} \in\, \mathbb{R}$.  So, as in the case of amplitude damping channel, symmetric initial state is necessary but not sufficient condition for the symmetry of state dynamics.  
Due to the form of dynamics of coherences, it is not simple to solve analytically conditions for symmetric entanglement decay for the whole family of X states. By using a restrictive assumption, $\rho_{23}\in\mathbb{R}$ or $\rho_{14}\in\mathbb{R}$, we can perform the analysis. For this subfamily of X states, entanglement decay is symmetric if and only if $\rho_{22} = \rho_{33}$ or $\rho_{11} = \rho_{44}$.

Interestingly, the families of initial states leading to symmetric state dynamics in bit-flip and bit-phase-flip channels are identical. Also for bit-phase-flip channel it is difficult to solve analytically, when entanglement decay is symmetric for the whole family of X states. 
If we set the same restriction as used for bit-flip channel above, we see that the necessary and sufficient conditions for symmetric entanglement decay in bit-phase-flip channel are the same as for bit-flip channel. On the other hand, by setting $\rho_{23},\,\rho_{14}\in\mathbb{C}\backslash\mathbb{R}$ in numerical tests, we could not find any initial states leading to symmetric entanglement decay for either of the channels. This serves as evidence for a claim that $\rho_{23}\in\mathbb{R}$ or $\rho_{14}\in\mathbb{R}$ is a necessary condition for an X state to have symmetric entanglement decay in these channels.

To conclude, as in Sec.~\ref{dep}--\ref{amp}, also for bit-flip and bit-phase-flip channels an asymmetric initial state with asymmetric state dynamics can lead to symmetric entanglement decay.

\section{Observations}
\begin{figure}[t]
\includegraphics[width=0.45\textwidth]{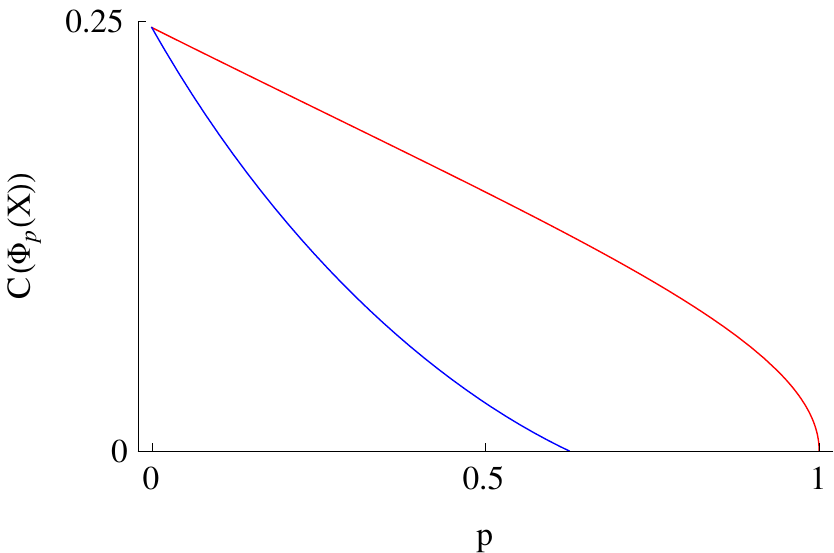}~~~
  \caption{
The red and blue curves correspond to situations with amplitude damping channel on qubit $U$ and $L$, respectively.
Influence of the noise on qubit $L$ leads to sudden death of entanglement, but when the noise affects the qubit $U$, entanglement decays asymptotically. We chose the input state as 
  $\rho_{11} = 0.35,\, 
\rho_{22} = 0.4,\, 
\rho_{33} = 0.05$ and $
\rho_{44} = 0.2$.
}\label{suddasymp}
\end{figure}
In Sec.~\ref{dep}--\ref{bfbpf} we studied families of initial states leading to symmetric and asymmetric decay of entanglement under different local channels. Next we concentrate on analyzing the implications of the results by using depolarizing and amplitude damping channels as examples. For plotting purposes, we set the coherence terms in each initial state to be maximal: $\vert\rho_{14}\vert = \sqrt{\rho_{11}\rho_{44}},\,\vert\rho_{23}\vert = \sqrt{\rho_{22}\rho_{33}}$.
\subsection{Asymmetry in sudden death of entanglement}
In Fig.~\ref{suddasymp} we present the concurrence of a state under amplitude damping channel. 
The plot shows that, for this choice of initial state, entanglement decay is sudden when amplitude damping noise affects qubit $L$, whereas entanglement decays asymptotically when the noise is acting on qubit $U$, instead. This means that, for this particular choice of initial state, the influence of the noise is significantly more harmful when qubit $L$ is affected by the noise.
\subsection{Entanglement decay as a resource}\label{ex2}
In this section we show that the asymmetry of entanglement decay can be used to gain information on one-sided channels.  A similar protocol was introduced in Ref.~\cite{pal}, where behavior of quantum discord and negativity were used to discriminate between channels.

\begin{figure}[t]
\centering
  \includegraphics[width=0.45\textwidth]{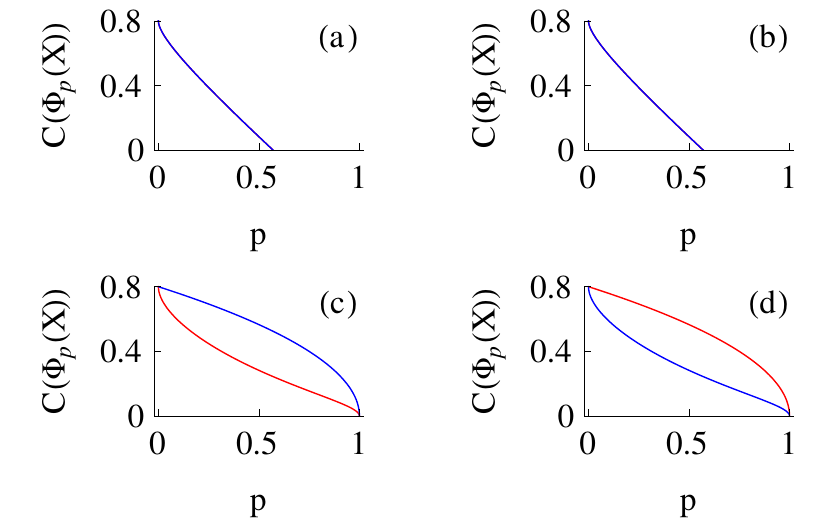}~~~
\caption{
Concurrence of an X state as a function of the channel strength parameter $p$ under depolarizing [(a)--(b)] and amplitude damping [(c)--(d)] channel. Here the red and blue curves correspond to the cases with local noise on qubit $U$ and $L$, respectively. 
In (a) and (c)  we have chosen 
$\rho_{11} = \rho_{44} = 0.4,\,
\rho_{22} = 0,\,$and $\rho_{33} = 0.2$. 
In (b) and (d) we have chosen 
$\rho_{11} = \rho_{44} = 0.4,\,
\rho_{22} = 0.2,\,$and 
$\rho_{33} = 0$.
}
\label{deplot1}
\end{figure}

In Fig.~\ref{deplot1}(a)--(b) we present the concurrence in depolarizing channel for two initial states. By choosing $\rho_{44} = \rho_{11}$ the noise is guaranteed to have identical influence independent of the location of the noise. The difference between the initial states used for Fig.~\ref{deplot1}(a) and \ref{deplot1}(b) is that the values of $\rho_{22}$ and $\rho_{33}$ were swapped. The form of Eq.~\eqref{depca} and \eqref{depcb} shows that, decay of entanglement is invariant under swapping $\rho_{22}$ and $\rho_{33}$, when $\rho_{44} = \rho_{11}$.

In Fig.~\ref{deplot1}(c) and \ref{deplot1}(d) we present the plots of concurrence in amplitude damping channel for initial states  used in Fig.~\ref{deplot1}(a) and \ref{deplot1}(b), respecticely. 
We see that $C\textbf{(}\Phi_p^U(X)\textbf{)} \neq C\textbf{(}\Phi_p^L(X)\textbf{)}$, for both initial states $X$, whenever $p\in(0,1)$. We notice that in Fig.~\ref{deplot1}(c) noise on qubit $U$ has more harmful effect on the concurrence and in \ref{deplot1}(d) the noise on qubit $L$ is more harmful. The swapping of the majorization of concurrence in Fig.~\ref{deplot1}(c) and \ref{deplot1}(d) can be seen directly from Eq.~\eqref{ampca} and \eqref{ampcb}. In fact, swapping $\rho_{22}$ and $\rho_{33}$ just swaps the curves corresponding to $C\textbf{(}\Phi_p^U(X)\textbf{)}$ and $C\textbf{(}\Phi_p^L(X)\textbf{)}$.

Plots (a)--(d) in Fig.~\ref{deplot1} show that, by using these two initial states, we can gain information of the channel which affects the system if we can assume that the noise is either depolarizing or amplitude damping type. If we let the noise influence one of the qubits and make a tomographic measurement\footnote{Due to the form of the fixed initial states, it is enough to determine five real valued parameters of the evolved density matrix, since none of the initially zero off-diagonal elements change under the influence of the channels.},  there are two possibilities, when using the two initial states presented above: either the concurrence has the same value for both initial states or one initial state leads to higher value of concurrence than the other. 
In the first case we know for sure that the noise was caused by a depolarizing channel and in the latter case the noise must have been amplitude damping. This means that this pair of initial states can be used to distinguish the two channels.

On the other hand, this can be done without knowing the value of $p$. So, after determining which channel affected the state, we can also obtain the value of $p$ for each channel by comparing the experimentally determined value of concurrence to the analytical solutions. 
Note that in the reasoning above we have not assumed anything about the location of the noise either. If we conclude that the noise was amplitude damping, we can compare the measured values of the concurrence for the two initial states. If the measured value of concurrence was smaller for the choice $\rho_{22} = 0$, then we know that the noise was affecting qubit $U$. On the other hand if the value of the concurrence is smaller for the choice $\rho_{33} = 0$, we know that the noise was influencing qubit $L$, instead. 

\begin{figure}[t]
\centering
  \includegraphics[width=0.45\textwidth]{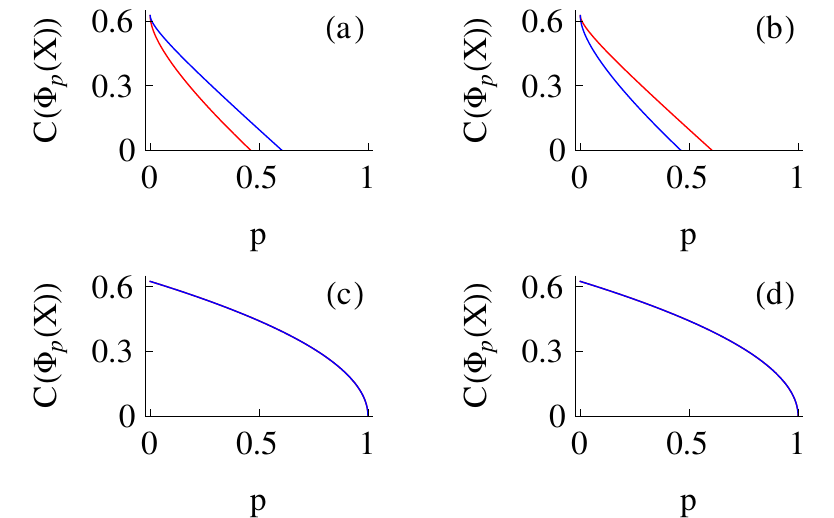}~~~
\caption{Concurrence of an X state as a function of the channel strength parameter $p$ under depolarizing [(a)--(b)] and amplitude damping [(c)--(d)] channel. Here the red and blue curves correspond to the cases with local noise on qubit $U$ and $L$, respectively. 
In (a) and (c) we have chosen 
$\rho_{11} = 0.2,\, 
\rho_{22} = 0.65,\, 
\rho_{33} = 0.15,
$ and 
$\rho_{44} = 0.$ 
In (b) and (d) we have chosen 
$\rho_{11} =0.2,\,
 \rho_{22}=0.15,\, 
 \rho_{33}=0.65,
 $ and 
 $\rho_{44} =0.$
Note that the situation is opposite to Fig.~\ref{deplot1}: now depolarizing channel causes asymmetric entanglement decay, and entanglement decays symmetrically in amplitude damping channel, instead.
}
\label{deplot2}
\end{figure}

Same reasoning can be done also for the case when the noise is depolarizing type. Plots and the corresponding pair of initial states are presented in Fig.~\ref{deplot2}. The protocol is not restricted just to  these two channels. By recalling the symmetry conditions of entanglement decay presented in Sec.~\ref{comb}--\ref{bfbpf}, one notices that we can now distinguish the set of channel combinations (a) and (c), the set of configurations (b) and (d), and individual amplitude damping, bit-flip and depolarizing channels.

\subsection{Entanglement decay and entropy}
In Ref.~\cite{hdecki} \.{Z}yczkowski $\&$ 
al.~studied asymmetric entanglement decay from the point of view of classical and quantum subsystems. They characterized a subsystem as classical if its von Neumann entropy is smaller than the von Neumann entropy of the total system, and as quantum if it is not classical. It was shown, through an example state, that when one of the subsystems is classical and the other one is quantum, noise affecting the classical subsystem decreases entanglement faster than if it was influencing the quantum subsystem.

\begin{figure}[t]
\centering
  \includegraphics[width=0.45\textwidth]{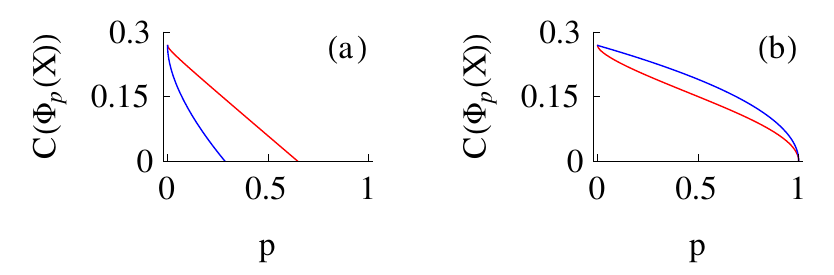}~~
\caption{Concurrence of a state as a function of the channel strength parameter $p$ for (a) amplitude damping and (b) depolarizing channel. Here the red and blue curves correspond to the cases with local noise on qubit $U$ and $L$, respectively. We have chosen 
$\rho_{11} = 0.9,\,
\rho_{22} = 
0,\, 
\rho_{33} = 0.08$,  
and
$\rho_{44} = 0.02$.}
\label{entr}
\end{figure}

In Fig.~\ref{entr} we present concurrence for another initial state as a function of channel strength parameter $p$ in amplitude damping and depolarizing channels. Von Neumann entropy of the initial total system state is $S(X) \approx 0.40$ and Von Neumann entropies of the reduced states of qubits $U$ and $L$ are $S(\text{tr}_L[X]) \approx 0.14$  and  $S(\text{tr}_U[X]) \approx 0.47$, respectively. Now subsystem $L$ is quantum but subsystem $U$ is classical. 
By looking at Fig.~\ref{entr} it is clear that the result of \cite{hdecki} does not hold generally, since in Fig.~\ref{entr}(a) entanglement decays faster when the noise affects qubit $L$ but in Fig.~\ref{entr}(b) the situation is the opposite. This also shows that, whether local noise on qubit $U$ results to faster or slower entanglement decay is not only a property of the initial state, but depends also on the channel.
\section{Decay of non-locality}
In their famous \color{black}paper \color{black} \cite{EPR}, Einstein, Podolsky and Rosen \color{black}concluded\color{black}, that some, hidden, variables should be added to quantum mechanics to restore locality to the theory. By assuming that the hidden variable theory is of local realistic nature, Clauser, Horne, Shimony and Holt derived the so-called CHSH inequality, which can be used to test the local hidden-variable theories \cite{CHSH}.

\color{black}The CHSH inequality can be written for a system in state $\rho$ as $\mathcal{F}\le 2$, where $\mathcal{F} =  \max_{\hat{a},\,\hat{a}',\,\hat{b},\,\hat{b}'}\big{\vert}\text{tr}[\rho(\hat{a}\otimes(\hat{b} + \hat{b}') + \hat{a}'\otimes(\hat{b} - \hat{b}'))]\big{\vert}$, is the Bell function, $\hat{a}$ and $\hat{a}'$ are some 
variables with values $\pm 1$ for qubit $U$ and $\hat{b}$ and $\hat{b}'$ are some 
variables with values $\pm 1$ for qubit $L$. Whenever $\mathcal{F}>2$, the locality assumption is violated, and we say that \textit{the state is non-local}.  CHSH inequality has been violated in experiments 
repeatedly, proving that the local hidden variable theories cannot be valid \cite{Hensen}.\color{black}

In \cite{bellvio} it was shown, that for 2-qubit states $\mathcal{F} = 2\sqrt{ u + \tilde{u}}$, where $u$ and $\tilde{u}$ are the two largest eigenvalues of $U_{\rho} = M^{\text{T}}_{\rho}M_{\rho}$, and $M$ is a matrix \color{black}defined by \color{black} $M_{i,j} = \text{tr}[\rho\sigma_i\otimes\sigma_j]$, where $i,\,j\in\{1,\,2,\,3\}$. For X states the eigenvalues of $U_{\rho}$ become
 \begin{equation}
\begin{aligned}
u_1 &= 4(\vert\rho_{14}\vert + \vert\rho_{23}\vert)^2\,,~~
u_2 = (r_{12} + r_{43})^2\,,\\ 
u_3 &= 4(\vert\rho_{14}\vert - \vert\rho_{23}\vert)^2\,.
\end{aligned}
\end{equation}
\color{black}Finally, \color{black}we get $\mathcal{F}^j =2\sqrt{\max \{\mathcal{F}_1^j, \mathcal{F}_2^j\}}$, where $\mathcal{F}_1^j = u_1^j + u_2^j$ and $\mathcal{F}_2^j = u_1^j + u_3^j$, and the superscript $j\in\{U, L\}$ tells whether the channel influences qubit $U$ or $L$.

Next we study the behavior of $\mathcal{F}$ under local depolarizing and amplitude damping channels \color{black}and their combinations\color{black}. As in \color{black}Sec.~\ref{concurr}\color{black}, also here we are interested in whether the dynamics of $\mathcal{F}$ depends on the location of the noise.
\subsection{Depolazing channel}
\begin{figure}[t]
\centering
  \includegraphics[width=0.45\textwidth]{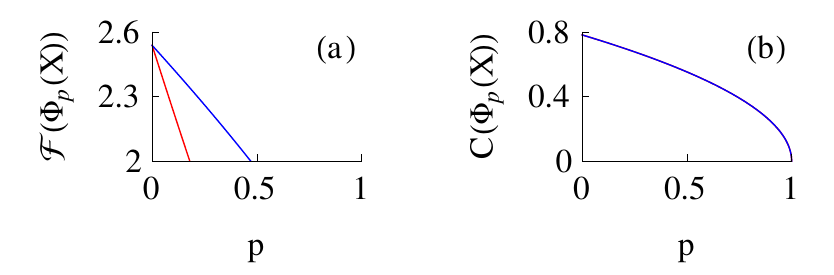}~~~
\caption{Behavior of (a) Bell function 
 and (b) concurrence as a function of channel parameter $p$.  
 In each plot the red and blue lines correspond to amplitude damping channel affecting qubits $U$ and $L$, respectively.  Here we have chosen  
 $\rho_{11} = \rho_{44} = 0,\, 
 \rho_{22} = 0.1875,\,$and $
 \rho_{33} = 0.8125$.
}
\label{nonlocalamp}
\end{figure}
For one-sided depolarizing channel, we see that 
\begin{equation}
\begin{aligned}
\mathcal{F}_1^U &= \mathcal{F}_1^L = 2q\sqrt{4 (\rho_{14} + \rho_{23})^2 + (r_{12} + r_{43})^2}\,,\\
\mathcal{F}_2^U &= \mathcal{F}_2^L =4q\sqrt{2(\vert\rho_{14}\vert^2 + \vert\rho_{23}\vert^2)}\,.
\end{aligned}
\end{equation}
This means that the effect of depolarizing noise on non-locality of the state is independent of the location for all X states. Since certain initial states lead to asymmetric entanglement decay in depolarizing channel, this implies that non-locality and entanglement behave in different way in terms of location of the noise.
\subsection{Amplitude damping channel}
For one-sided amplitude damping channel, we see that 
\begin{align}\label{BA}
\mathcal{F}_1^U = & 2\sqrt{4q(\vert\rho_{14}\vert + \vert\rho_{23}\vert)^2 
+ \big[r_{12} + (2p - 1)r_{34}\big]^2}\,,\\ 
\mathcal{F}_2^U = &~ \mathcal{F}_2^L = 4\sqrt{2q(\vert\rho_{14}\vert^2 + \vert\rho_{23}\vert^2)}\,,\\
\label{BB}
\mathcal{F}_1^L = &2\sqrt{4q(\vert\rho_{14}\vert + \vert\rho_{23}\vert)^2
+ \big[r_{13} + (2p - 1)r_{24}\big]^2}\,.
\end{align}
Comparison of Eq.~\eqref{BA} and \eqref{BB} shows that now the location of the noise makes a difference, unlike in the case of depolarizing channel. We see that $\mathcal{F}_1^U = \mathcal{F}_1^L\,\, \forall p\in[0,1]$ if and only if $\rho_{22} = \rho_{33}$ or $\rho_{11} = \rho_{44} = (\rho_{22} + \rho_{33} )/2 = 1/4 $. The first condition is satisfied by all symmetric initial states, so, as in the case of concurrence, also symmetric decay of non-locality seems to be a direct consequence of symmetric initial state. On the other hand, the second condition does not require symmetric initial state. For all initial X states satisfying the second condition, the value of $\mathcal{F}$ is maximized with the choice $\rho_{11} = \rho_{22} = \rho_{33} = \rho_{44} = \vert\rho_{14}\vert = \vert\rho_{23}\vert = 1/4 $. For this state we get $\mathcal{F} = 2$, which is not interesting in the context of decay of non-locality, since such state is initially local.

In Fig.~\ref{nonlocalamp} we present the behavior of \color{black}$\mathcal{F}$ \color{black} in amplitude damping channel. \color{black}The plot illustrates \color{black} the difference between the two noise locations: when the channel affects qubit $U$, the decay is twice as fast compared to the case of noise on qubit $L$. For comparison, we present also the plot of concurrence for the same initial state. In contrast to what happens to the Bell function, concurrence decays independently of the location of the noise.

\begin{table}
\centering
\caption{Behavior of entanglement decay (ED) and decay of non-locality (ND) for different families of initial states.}
\label{deptab}
\textbf{(a)} \textit{Depolarizing channel.}\\
\begin{tabular}{c|c|c|c|c}
\hline\hline
	& $ \rho_{22} = \rho_{33} $ & $\rho_{22} \neq \rho_{33}\,\,\&$    & $\rho_{22} \neq \rho_{33}\,\,\&$     & $\rho_{22} \neq \rho_{33}\,\,\&$                                                
	\\
	&                           & $\rho_{11} = \rho_{44} \neq 0$ & $\rho_{11} = \rho_{44} = 0 $ & $\rho_{11} \neq \rho_{44} = 0 $ 
	\\ \hline
ED     & Symmetric                 & Symmetric                      & Symmetric                       & \textbf{Asymmetric}                         
	\\ 
ND & Symmetric                 & Symmetric                      & Symmetric                       & \textbf{Symmetric}                          
	\\ \hline\hline
\end{tabular}\\
\bigskip
\textbf{(b)} \textit{Amplitude damping channel.}\\
\begin{tabular}{c|c|c|c|c}
\hline\hline
                             & $ \rho_{22} = \rho_{33} $ 	& $\rho_{22} \neq \rho_{33}\,\,\&$   		& $\rho_{22} \neq \rho_{33}\,\,\&$	& $\rho_{22} \neq \rho_{33}\,\,\&$					
                             \\
                             &                         				& $\rho_{11} = \rho_{44} \neq 0$ 	& $\rho_{11} = \rho_{44} = 0 $ 	& $\rho_{11} \neq \rho_{44} = 0 $ 	
                             \\ \hline
ED     		& Symmetric			& Asymmetric			& \textbf{Symmetric}			& \textbf{Symmetric}				
\\
ND 		& Symmetric			& Asymmetric 			& \textbf{Asymmetric}			& \textbf{Asymmetric}			
\\ \hline\hline
\end{tabular}
\end{table}
We have gethered in Tab.~\ref{deptab} the families of initial states leading to interesting dynamics for concurrence and Bell function. 
The most interesting result is obtained with choices $ \rho_{22} \neq \rho_{33},\, \rho_{11} \neq \rho_{44} = 0$. For this family, depolarizing channel leads to asymmetric entanglement decay and symmetric decay of Bell function. Contrary to this, the result for amplitude damping channel is the opposite: symmetric entanglement decay and asymmetric decay of Bell function. This means that there is no hierarchy between the asymmetry of entanglement decay and decay of Bell function: asymmetry of one property does not imply or exclude the asymmetry of the other.

\subsection{Channel combinations}

To complete the study of Bell function decay under depolarizing and amplitude damping channels, we perform the analysis on combinations presented in Sec.~\ref{comb}. Trivially again \color{black}$\mathcal{F}$ \color{black} decays symmetrically in (a) if both channels are the same. Then again, if they are different, the decay is symmetric if and only if $\rho_{22} = \rho_{33}$. In fact, this is the same condition as for symmetric entanglement decay in this configuration.  If in (b), (c), or (d) all channels are depolarizing (amplitude damping) type, the symmetry conditions for \color{black}$\mathcal{F}$ \color{black} decay are the same as for single depolarizing (amplitude damping) channel. So, as in the case of entanglement decay, also the symmetry of Bell function decay seems to be invariant under repetition of the same local noise.

On the other hand, if in combinations (b) -- (d) there is at least one copy of each channel, \color{black}$\mathcal{F}$ \color{black} decays symmetrically if and only if $\rho_{22} = \rho_{33}$. We note, that this differs from the conditions of symmetric entanglement decay in this configuration. The only exception appears in configuration (c). If $\Phi$ and $\Xi$ are amplitude damping channels and $\Psi$ is depolarizing channel, \color{black}$\mathcal{F}$ \color{black} decays symmetrically for all initial X states. 

For the sake of example, let us assume a situation, in which symmetric decay of non-locality is desired. Configuration (a) can be divided into two cases in terms of symmetry conditions: $\Phi  = \Psi$, leading always to symmetric decay of \color{black}$\mathcal{F}$, \color{black} and $\Phi  \neq \Psi$ leading  to symmetric decay of \color{black}$\mathcal{F}$ \color{black}  if and only if $\rho_{22} = \rho_{33}$. In the latter case we can now achieve symmetry for all initial X states, by adding amplitude damping noise before depolarizing channel. It is worth noting, that there is something special about this configuration, since it is impossible to induce the symmetry by adding amplitude damping noise after the depolarizing channel or on the same side with the original amplitude damping channel. Also, if the original configuration has just a single amplitude damping channel, one can achieve symmetry by adding local amplitude damping and depolarizing noises on the qubits.

The analysis in Sec.~\ref{comb} shows, that such phenomenon does not occur for entanglement decay in simple combinations of depolarizing and amplitude damping channels: in the case described above, adding one amplitude damping \color{black}or depolarizing \color{black} channel in any possible location has no effect in the symmetry of entanglement decay.

\section{Decay of trace distance discord}

Since all quantum correlations cannot be described by entanglement and non-locality, we conclude our study by considering the dynamics of a more general type of correlation, \textit{quantum discord}.
Due to the difficulty of computing and comparing the exact values of quantum discord, geometric measures have been developed. Geometric discord measures are based on the smallest distance between the given state $\rho$ and the set of states with zero discord. A state $\tilde{\rho}$ has zero discord if and only if it can be decomposed as
\begin{align}
\tilde{\rho} = \sum_{j}\ket{\alpha_j}\bra{\alpha_j}\otimes\rho^L(j)\,,
\end{align}
where $\{\ket{\alpha_j}\}_j$ is a set of orthogonal vectors in $\mathcal{H}^U$ and $\rho^L(j)$ are positive operators in $\mathcal{H}^L$. The choice of metric used to measure the distance determines the properties of geometric discord. A good choice for metric is trace distance $D_{\text{tr}}(\rho,\,\xi) = \vert\vert\rho - \xi\vert\vert_{\text{tr}}$, where $\vert\vert\rho\vert\vert_{\text{tr}} = \text{tr}[\sqrt{\rho^{\dagger}\rho}]/2$ is the trace norm. With this norm, \textit{the trace distance discord} can be defined as
\begin{align}
D(\rho) = \min_{\tilde{\rho}}D_{\text{tr}}(\rho,\,\tilde{\rho})\,,
\end{align}
where the minimization is taken over the set all states $\tilde{\rho}$ with zero discord.

In Ref.~\cite{trdisc} it was shown, that for an arbitrary X state the trace distance discord can be calculated as
\begin{align}\label{DX}
D(X) &= \frac{1}{2}\sqrt{\frac{\gamma_1^2\max\{\gamma_3^2,\,\gamma_2^2+x^2\}-\gamma_2^2\min\{\gamma_3^2,\,\gamma_1^2\}}{\max\{\gamma_3^2,\,\gamma_2^2+x^2\}-\min\{\gamma_3^2,\gamma_1^2\}+\gamma_1^2-\gamma_2^2}}\,,\\
\gamma_{1} &= 2(\rho_{32} + \rho_{41}),\,~~~~~
\gamma_{2} = 2(\rho_{32} - \rho_{41}),\,\\ 
\gamma_{3} &= 1 - 2(\rho_{22} + \rho_{33}),\,~
x = 2(\rho_{11} + \rho_{22}) - 1\,.
\end{align}
Assuming $\rho_{23},\,\rho_{14}\in\mathbb{R}$ simplifies Eq.~\eqref{DX} into
\begin{align}
\label{g1}
D(X) &= \frac{\vert\gamma_1\vert}{2},\,~~\text{when}~~\vert\gamma_3\vert \ge \vert\gamma_1\vert\\
\label{g3}
D(X) &= \frac{\vert\gamma_3\vert}{2},\,~~\text{when}~~\vert\gamma_3\vert < \vert\gamma_1\vert\,\, \&\,\, \gamma^2_3 \ge \gamma_2^2 + x^2\\
\label{gall}
D(X) &= \frac{1}{2}\sqrt{\frac{\gamma^2_1(\gamma^2_2 + x^2) - \gamma^2_2\gamma^2_3}{\gamma^2_1 - \gamma^2_3 + x^2}},\,~~~~\text{ otherwise.}
\end{align}

\subsection{Asymmetric discord dynamics}

In depolarizing, bit-flip and bit-phase-flip channels the parameters $\gamma_1,\,\gamma_2$ and $\gamma_3$ evolve symmetrically with respect to noise location. This means that whenever the conditions of  Eq.~\eqref{g1} or Eq.~\eqref{g3} are satisfied throughout the dynamics, trace distance discord behaves symmetrically. On the other hand, we see that in the case of Eq.~\eqref{gall}, discord evolves symmetrically if and only if the parameter  $x$ has symmetric dynamics. This is equivalent to using initial state with $x = 0$. In the case of Eq.~\eqref{gall}, this \color{black}choice \color{black} leads to $D(X) = \vert\gamma_2\vert/2$. 

\begin{figure}[t]
  \centering
  \def\svgwidth{225pt}
  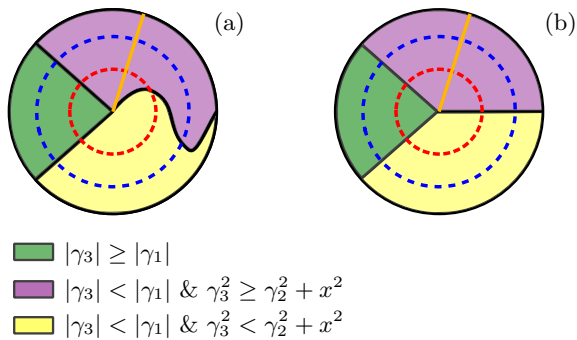
    \caption{\color{black}Visualization for possible trajectories of state dynamics (color online). The cases are divided between initial states satisfying (a) $x \neq 0$ and (b) $x = 0$. The space of states is represented by a clock-face and split into three colored  segments. The channel strength $p$ corresponds to the rotation angle $\theta(p)$ of the orange hand of the clock. The red and blue dashed circles represent possible trajectories of $\Phi^U_p(X)$ and $\Phi^L_p(X)$, respectively. As $p$ increases, the clock hand rotates moving the evolved state along the dashed circle which corresponds to the noise location. The value of $\theta(0)$ is determined by the initial state and the behavior of $\theta(p)$ depends on the channel $\Phi_p$ and the initial state. \color{black}
    }
  \label{discdyn}
\end{figure}
  
\color{black}Two exemplary cases of states evolving in subspaces of X states, defined by conditions in Eq.~\eqref{g1} -- \eqref{gall}, are illustrated in Fig.~\ref{discdyn}.
In Fig.~\ref{discdyn}(a) we see, that noise on qubit $U$ makes the state enter or exit the green segment if and only if noise on qubit $L$ does so. Contrary to this, the line between purple and yellow segments is crossed by noise on one qubit when noise on the other one does not make the state cross the line. Instead in Fig.~\ref{discdyn}(b), each segment boarder is crossed by noise on one qubit if and only if noise on the other one makes the state cross the line. This case is special also because decay of discord is symmetric in all sections, and thus through the whole dynamics.

The non-trivial cases occur when noise on one qubit causes the evolved state $\Phi_p(X)$ to satisfy $\vert\gamma_3\vert < \vert\gamma_1\vert\,\, \&\,\, \gamma^2_3 \ge \gamma_2^2 + x^2$ but noise on the other one leads to a state satisfying $\vert\gamma_3\vert < \vert\gamma_1\vert\,\, \&\,\, \gamma^2_3 < \gamma_2^2 + x^2$, instead. This situation is illustrated in Fig.~\ref{discdyn} (a) as the curved boarder between the purple and yellow segments. For depolarizing, bit-flip and bit-phase-flip channels this case never results to symmetric decay of discord. Thus we conclude, that symmetry of discord dynamics requires that both local channels map the state into the same segment of the state space at each value of $p$.\color{black}

For depolarizing and bit-phase-flip channels, initial state satisfying $\vert\gamma_3\vert \ge \vert\gamma_1\vert$ guarantees that the property is preserved throughout the dynamics. In the sense of Fig.~\ref{discdyn} this means that any initial state inside the green segment never exits it during dynamics. Neither can any state outside the region enter it. As an interesting example, we study the maximally discordant mixed two-qubit states (MDM). In \cite{mdms} the analytic form of MDM's was solved. They are all  X states and it is easy to see that they satisfy $\vert\gamma_3\vert\ge\vert\gamma_1\vert$, but their state dynamics is symmetric only in special cases. Thus all two-qubit MDM's have symmetric discord dynamics under depolarizing and bit-phase-flip channel.

For bit-flip channel the conditions are more restrictive. If initial state satisfies $\vert\gamma_3\vert \ge \vert\gamma_1\vert$, the evolved state $\Phi_p(X)$ satisfies it also if and only if $p \le 1/2 - \vert\rho_{23} + \rho_{14}\vert/\vert1 - 2(\rho_{22} + \rho_{33})\vert$ or $p \ge 1/2 + \vert\rho_{23} + \rho_{14}\vert/\vert1 - 2(\rho_{22} + \rho_{33})\vert$. We note, that the values of $p$ keeping the evolved state $\Phi_p(X)$ inside the region are determined by $D(X)$. Especially, $\Phi_p(X)$ satisfies $\vert\gamma_3\vert \ge \vert\gamma_1\vert~ \forall p\in[0,1]$ if and only if $D(X) = \vert\rho_{23} + \rho_{14}\vert = 0$.
\color{black}

 If we assume $\vert\gamma_3\vert < \vert\gamma_1\vert$ instead, decay of discord is symmetric for all of the channels, whenever $\gamma^2_3 \ge \gamma_2^2 + x^2$ is satisfied \color{black}by \color{black} both $\Phi^U_p(X)$ and $\Phi^L_p(X)$. On the other hand, if $\gamma^2_3 < \gamma_2^2 + x^2$ for both $\Phi_p^U(X)$ and $\Phi_p^L(X)$, decay of discord is symmetric for these channels if and only if $\rho_{11} + \rho_{22} = 1/2$. 

Now we assume that $\vert\gamma_3\vert < \vert\gamma_1\vert$ and study, which initial states stay inside or exit the region defined by $\gamma^2_3 \ge \gamma_2^2 + x^2$ under different local noises. For bit-flip or depolarizing noise on qubit $U$, no initial state inside the region exits it nor does any initial state outside the region enter it. Contrary to this, under bit-flip or depolarizing noise on qubit $L$, an initial state stays inside the region if and only if 
$x = 0$. This means, that initial states violating $x = 0$ are mapped into region $\gamma^2_3 < \gamma_2^2 + x^2$ by noises on qubit $L$,   and thus they have asymmetric decay of discord.

For bit-phase-flip channel, the set of states staying inside the region is more exclusive. Noise on qubit $U$ preserves the property if and only if the initial state satisfies  $\gamma_{2} = 0$. Bit-phase-flip noise on qubit $L$ keeps the state inside the region if and only if $\gamma_{2} = x = 0$. Like in the case of the other channels, also here we see that initial states, satisfying $\gamma_{2} = 0$ but violating $x = 0$, stay inside the region when noise affects qubit $U$ but exit it when noise is applied on qubit $L$ instead,  \color{black}resulting to asymmetric discord dynamics\color{black}.

To conclude, we see that for all of the channels and all three segments there exist well-defined families of states which stay within the segment they started from. There is no preference in the direction of crossing the segment boundaries: each segment has initial states exiting it and initial states from other segments entering it during the evolution. The only exception is segment $\vert\gamma_3\vert \ge \vert\gamma_1\vert$ in depolarizing and bit-phase-flip channels: no initial state outside the segment can enter it nor does any state initially inside exit it. 

\subsection{Time-invariant quantum discord}

\color{black}
Finally, we study the so-called \textit{time-invariant discord} phenomenon, where the value of discord is not influenced by the channel.
Local depolarizing noises never induce time-invariant discord. Instead bit-phase-flip and bit-flip channels do. Discord is invariant under local bit-phase-flip channel if and only if $x = 0$ and $\vert\gamma_3\vert<\vert\gamma_2\vert$. First of the conditions is preserved in the dynamics for all initial states satisfying it and the second one holds for the evolved state $\Phi_p^U(X)$ if and only if $\Phi_p^L(X)$ satisfies it. So we see that time-invariant discord occurs under bit-phase-flip channel only if the discord dynamics is symmetric.

For a bit-flip channel, there are two disjoint families of states leading to time-invariant discord: initial states satisfying $\vert\gamma_3\vert\ge\vert\gamma_1\vert>0$, or $\vert\gamma_3\vert = 0$  and $\rho_{23} = \rho_{14}\neq0$. All initial states satisfying either of the conditions satisfies the condition through the whole dynamics. States in the first family have symmetric discord dynamics.  Instead, for initial states in the latter family, discord dynamics is asymmetric. In fact the time-invariant discord occurs only when the local noise is applied on qubit $L$. If the noise affects qubit $U$ instead, the value of discord goes to zero for all initial states of the family.\color{black}

\section{Summary and discussion}
\color{black}We have studied the dynamics of concurrence, Bell function and trace distance discord under one-qubit channels and their combinations. The channels we considered were depolarizing, amplitude damping, bit-flip, and bit-phase-flip channels and simple combinations of depolarizing and amplitude damping channels.
We saw that even though the input state, or even the dynamics of the state, is asymmetric, entanglement, non-locality and discord can decay symmetrically. We noticed that the families of asymmetric states leading to symmetric entanglement or Bell function decay are not the same for different channels or channel combinations. 
Thus, by measuring how concurrence or Bell function decays, one can deduce which noise was affecting the two-qubit system, which qubit it affected, and how large is the channel strength parameter $p$. 

We also saw that, for some initial states, entanglement decay is sudden when the noise affects one qubit and asymptotic when the noise influences the other. Also, the same initial state can lead to sudden death of entanglement for one type of channel and asymptotic decay for another. For one-sided amplitude damping channel, initial states with symmetric entanglement decay can lead to asymmetric decay of Bell function and the opposite happens in depolarizing channel. This means that there is no natural hierarchy between the asymmetric decays of concurrence and non-locality. 

Finally, we studied the dynamics of trace distance discord noticing that the total space of states can be divided into three disjoint regions. We saw that symmetric decay of discord requires that local noise affecting the system maps the initial state into the same region at the same value of parameter $p$. We characterized families of initial states leading to symmetric and asymmetric discord dynamics for each channel. We also characterized families of initial states leading to time-invariant discord and saw that for some of these families discord dynamics is symmetric but for one family time-invariant discord occurs only when noise affects qubit $L$.

\section*{Acknowledgements}
The authors wish to thank Gerardo Adesso for providing helpful discussions. The Turku group acknowledges the financial support from Horizon 2020 EU collaborative project QuProCS (Grant Agreement 641277), Magnus Ehrnrooth Foundation, and the Academy of Finland (Project no. 287750). 
G.K. is grateful to Sao Paulo Research Foundation (FAPESP) for the support given under the grant number 2012/18558-5. 
The Hefei group was supported by the National Natural Science Foundation of China (Nos. 61327901, 11274289, 11325419), the Strategic Priority Research Program (B) of the Chinese Academy of Sciences (Grant No. XDB01030300), Key Research Program of Frontier Sciences, CAS (No. QYZDY-SSW-SLH003).

\begin{widetext}
\begin{appendix}
\section{Bit-flip channel}\label{bf}
Corresponding to Eq.~\eqref{totkrausa} and \eqref{totkrausb}, the dynamics of an arbitrary X state under one-sided bit-flip channels can be written as
 \begin{equation}
\begin{aligned}
\Phi_p^U(X)&=
\begin{pmatrix}
q \rho_{11} + p\rho_{33}			&0									&0									&q\rho_{14} + p\rho_{23}^*	\\
0									&q\rho_{22} + \rho_{44}p		&q\rho_{23} + p\rho_{14}^*		&0	\\
0									&q\rho_{23}^* + p\rho_{14}		&q\rho_{33} + \rho_{11}p		&0	\\
q\rho_{14}^*  + p\rho_{23}		&0									&0									&q\rho_{44} + \rho_{22}p
\end{pmatrix},\,
\\
\Phi_p^L(X)&=
\begin{pmatrix}
q \rho_{11} + p\rho_{22}			&0									&0									&q\rho_{14} + p\rho_{23}	\\
0									&q\rho_{22} + \rho_{11}p		&q\rho_{23} + p\rho_{14}		&0	\\
0									&q\rho_{23}^* + p\rho_{14}^*	&q\rho_{33} + \rho_{44}p		&0	\\
q\rho_{14}^*  + p\rho_{23}^*	&0									&0									&q\rho_{44} + \rho_{33}p	
\end{pmatrix}
.
\end{aligned}
\end{equation}
By using Eq.~\eqref{conc} we get as the concurrences of the output states
\begin{equation}
\begin{aligned}
&C\textbf{(}\Phi_p^U(X)\textbf{)}=2\max\Big{\{}0,\, \vert \rho_{14}p + \rho_{23}^*q\vert - \sqrt{(\rho_{22}p + \rho_{44}q)(\rho_{33}p + \rho_{11}q)}
,\,
 \vert \rho_{14}^*q + \rho_{23}p\vert - \sqrt{(\rho_{11}p + \rho_{33}q)(\rho_{44}p + \rho_{22}q)}\Big{\}},\\
&C\textbf{(}\Phi_p^L(X)\textbf{)}=2\max\Big{\{}0,\,\vert \rho_{14}^*p + \rho_{23}^*q\vert - \sqrt{(\rho_{33}p + \rho_{44}q)(\rho_{22}p + \rho_{11}q)}
,\,
 \vert \rho_{14}^*q + \rho_{23}^*p\vert - \sqrt{(\rho_{11}p + \rho_{22}q)(\rho_{44}p + \rho_{33}q)}\Big{\}}.
\end{aligned}
\end{equation}
Due to the form of the solutions it is not as simple to solve analytically, when the decay of entanglement is symmetric. Still we can find a set of sufficient conditions for symmetric entanglement decay:
\begin{align}
(\rho_{14}\in\mathbb{R}\,\, \text{or}\,\,\rho_{23}\in\mathbb{R})\,\, \text{and}\,\, (\rho_{22} = \rho_{33}\,\, \text{or}\,\,\rho_{11} = \rho_{44})
\end{align}
\section{Bit-phase-flip channel}\label{bpf}
The dynamics of an arbitrary X state under one-sided bit-phase-flip channels can be written as
\begin{equation}
\begin{aligned}
\Phi_p^U(X)&=
\begin{pmatrix}
q \rho_{11} + p\rho_{33}		&0								&0								&q\rho_{14} - p\rho_{23}^*	\\
0								&q\rho_{22} + \rho_{44}p	&q\rho_{23} - p\rho_{14}^*	&0	\\
0								&q\rho_{23}^* - p\rho_{14}	&q\rho_{33} + \rho_{11}p	&0	\\
q\rho_{14}^*  - p\rho_{23}	&0								&0								&q\rho_{44} + \rho_{22}p	
\end{pmatrix},
\\
\Phi_p^L(X)&=
\begin{pmatrix}
q \rho_{11} + p\rho_{22}			&0										&0								&q\rho_{14} - p\rho_{23}	\\
0									&q\rho_{22} + \rho_{11}p			&q\rho_{23} - p\rho_{14}		&0	\\
0									&q\rho_{23}^* - p\rho_{14}^*		&q\rho_{33} + \rho_{44}p	&0	\\
q\rho_{14}^*  - p\rho_{23}^*	&0										&0								&q\rho_{44} + \rho_{33}p	
\end{pmatrix}
.
\end{aligned}
\end{equation}
By using Eq.~\eqref{conc} we get as the concurrences of the output states
\begin{equation}
\begin{aligned}
&C\textbf{(}\Phi_p^U(X)\textbf{)}=2\max\Big{\{}0,\, \vert  \rho_{23}^*q-\rho_{14}p\vert - \sqrt{(\rho_{22}p + \rho_{44}q)(\rho_{33}p + \rho_{11}q)}
,\,
 \vert \rho_{14}^*q + \rho_{23}p\vert - \sqrt{(\rho_{11}p + \rho_{33}q)(\rho_{44}p + \rho_{22}q)}\Big{\}},\\
&C\textbf{(}\Phi_p^L(X)\textbf{)}=2\max\Big{\{}0,\,\vert \rho_{23}^*q  -\rho_{14}^*p \vert - \sqrt{(\rho_{33}p + \rho_{44}q)(\rho_{22}p + \rho_{11}q)}
,\,
 \vert \rho_{14}^*q - \rho_{23}^*p\vert - \sqrt{(\rho_{11}p + \rho_{22}q)(\rho_{44}p + \rho_{33}q)}\Big{\}}.
\end{aligned}
\end{equation}

\end{appendix}
\end{widetext}

\end{document}